\documentclass[review,number,sort&compress]{elsarticle}

\usepackage{lineno}
\usepackage{hyperref}
\usepackage{comment}
\usepackage{graphicx}
\usepackage{subcaption}
\usepackage{capt-of}
\usepackage{color}
\usepackage{siunitx}
\usepackage{textcomp}
\usepackage{todonotes}
\usepackage{float}

\DeclareSIUnit[number-unit-product = ]\percent{\char`\%}
\usepackage[nowatermark]{fixmetodonotes}

\journal{Nuclear Instruments and Methods A}

\begin{document}

\begin{frontmatter}

\title{Water-based Liquid Scintillator Detector as a New Technology Testbed for Neutrino Studies in Turkey}


\author[Davis]{Vincent Fischer}

\author[ISU]{Emrah Tiras\corref{mycorrespondingauthor}}
\cortext[mycorrespondingauthor]{Corresponding author}
\ead{etiras@fnal.gov}

\address[Davis]{Department of Physics, University of California, Davis, CA 95616, USA}
\address[ISU]{Department of Physics and Astronomy, Iowa State University, Ames, IA 50011, USA}

\begin{abstract}
This study investigates the deployment of a medium-scale neutrino detector near Turkey's first nuclear power plant, the Akkuyu Nuclear Power Plant. 
The aim of this detector is to become a modular testbed for new technologies in the fields of new detection media and innovative photosensors. 
Such technologies include Water-based Liquid Scintillator (WbLS), Large Area Picosecond Photo-Detectors (LAPPDs), dichroic Winston cones, and large area silicon photomultiplier modules. 
The detector could be used for instantaneous monitoring of the Akkuyu Nuclear Power Plant via its antineutrino flux. 
In addition to its physics and technological goals, it would be an invaluable opportunity for the nuclear and particle physics community in Turkey to play a role in the development of next generation of particle detectors in the field of neutrino physics. 
\\
\end{abstract}


\end{frontmatter}

\newpage 
\section{Introduction}
\label{sec:intro}
Neutrinos, the most abundant massive particles in the universe, are believed to be vital ingredients in many mysterious physics processes. 
In addition to the neutrino mass being an experimental palpable evidence beyond the Standard Model of particle physics, there are many questions related to neutrinos still need to be answered. 

Neutrinos are generated by numerous sources, cosmic rays, the Sun, supernovae, nuclear decay, nuclear reactors and particle accelerators. 
They can provide information about fusion processes that power the Sun and stars, about radioactive decays inside the Earth, or the fission reactions utilized in nuclear reactors. 
The Akkuyu Nuclear Power Plant (Akkuyu NPP), which will be the first to be built in Turkey, offers a great opportunity for building the country's very first neutrino detector and fuelling the expanding field of neutrino science in Turkey.  

In their recent paper, S. Ozturk \text{et. al.}~\cite{Ozturk:2016tjn} proposed a 1-ton water Cherenkov detector doped with gadolinium (Gd) and instrumented with 10-inch photomultiplier tubes (PMTs) as a way to monitor the Akkuyu NPP at very short distances. Another paper by M. Kandemir and A. Cakir~\cite{Kandemir:2019dfs} proposed a mobile near-field ($<$100 m) detector based on plastic scintillator bars and 3-inch PMTs. 
More recently, S. Ozturk has conducted some simulation studies for another short-range detector with gadolinium-loaded plastic scintillator bars and PMTs using multivariate analysis techniques~\cite{Ozturk:2019bul}. 

All of those studies propose near-field, small size detectors likely on the site of the Akkuyu NPP. By contrast, the study presented here aims for an international collaborative effort to build an innovative medium-size neutrino detector a few kilometers away from the site of the power plant. The project goals will be: 

\begin{itemize}
    \item monitoring the Akkuyu NPP in a non-intrusive manner by measuring the antineutrino flux originating from its reactors, 
    \item deploying and testing new technologies (Gd-loaded WbLS~\cite{WBLS1}, LAPPDs~\cite{Adams:2016tfm}, dichroic Winston cones, also nicknamed dichroicons~\cite{Kaptanoglu:2018sus,Kaptanoglu:2019gtg}, etc.) for future neutrino experiments such as WATCHMAN~\cite{Askins:2015bmb} and THEIA~\cite{Askins:2019oqj},
    \item recruiting and training the next generation of detector experts and neutrino physicists in Turkey.  
\end{itemize}

In the meantime, the aforementioned near-field, small-size neutrino detectors may fall within this collaborative effort and could be supported. 
Our aim is to develop a comprehensive Research \& Development (R\&D) program based around Neutrino Physics but that will benefit many related fields including High Energy Physics (HEP), Nuclear Physics (NP) and Astroparticle Physics (AP). Researchers from each of these disciplines would be invited to join this collaborative effort to reach its goals in the field of neutrinos. International collaborations with groups wishing to deploy and tests new technologies are also foreseen.

This paper describes a detector for monitoring the Akkuyu NPP that incorporates several novel and growing technologies:

\begin{itemize}
    \item the use of Water-based Liquid Scintillator (WbLS), a new detection medium that combines the tracking capabilities of Cherenkov detectors with the energy resolution and sensitivity of scintillation detectors,
    \item Gd-loaded media to enable neutron tagging, improving detection efficiency and reconstruction capabilities~\cite{Beacom:2003nk},
    \item multiple new photodetector technologies that enable the separation of light components and would allow multi-track reconstruction and the identification of more exclusive final states. 
\end{itemize}

We consider a 30-ton WbLS neutrino detector that would reside in an underground laboratory below one of the hills in the immediate vicinity (1-3~km) of the Akkuyu NPP. 

\section{New Technologies}
\label{sec:new_tech}

The increasing need for precise kinematic reconstruction of particle interactions is driving a global effort in the particle physics community to develop new technologies in the fields of detection media and photon detectors.
The following section presents a non-exhaustive list of active research areas that are of interest for a future testbed in Turkey with immediate application in a neutrino experiment. 

\subsection{Water-based Liquid Scintillator}
\label{subsec:wbls}

Water-based Liquid Scintillator (WbLS) is a homogeneous mixture of pure water and organic liquid scintillator developed at Brookhaven National Laboratory~\cite{WBLS1}.
The addition of a high light yield liquid scintillator into water allows the detection of particles with an energy below the Cherenkov threshold while preserving the advantages of pure water; directional capability, cost and scalability.
By adjusting the content of liquid scintillator in the mixture, one can maximize the potential of a WbLS detector.
Previous successful attempts to load metals in water as well as in liquid scintillator are a good indicator that WbLS can also be loaded with a neutron detection enhancing isotope such as gadolinium (Gd) or lithium (Li).

An active worldwide R\&D program is currently ongoing to fully develop and characterize WbLS and future large-scale projects such as THEIA have already chosen it as their main target medium.
Other smaller projects such as ANNIE~\cite{Back:2017kfo} and WATCHMAN also plan to deploy WbLS as a detection medium.

For the purpose of this study, we made only two realistic assumptions concerning WbLS performance: (1) WbLS can be loaded with an element such as Gd or Li to enhance its neutron detection capabilities, and (2) a WbLS mixture with a 10\% liquid scintillator content can be manufactured and utilized.

\subsection{New Photosensors}
\label{subsec:photosensors}

\subsubsection{Large Area Picosecond Photo-Detectors}

Large Area Picosecond Photo-Detectors (LAPPDs) are 20~cm x 20~cm microchannel plate based photosensors~\cite{Adams:2016tfm}. 
Their thin form factor and segmented anode strip readout allows them to reach a single photoelectron time resolution of about 50~ps and sub-cm spatial resolution~\cite{Lyashenko:2019tdj}.
These two characteristics make them an invaluable tool not just in detecting single photons but also reconstructing temporal and spatial light patterns such as Cherenkov rings.
LAPPDs have been tested on dedicated setups in the past~\cite{Tiras2019} and will be deployed and operated in the ANNIE experiment with the goal of demonstrating their capabilities as an imaging detector for particle physics.
LAPPDs in combination with WbLS could improve the separation between directional Cherenkov light and isotropic scintillation light allowing for accurate kinematic reconstruction in addition to precise calorimetric measurement.

\subsubsection{Wavelength-selective Light Concentrators}

Scintillation and Cherenkov light exhibit different wavelength spectra.
While Cherenkov light has a broad distribution with an intensity decreasing as the wavelength increases, scintillation light is typically narrower and centered around 360~nm depending on the scintillating fluor.
Performing a spectral filtration using dichroic filters allows separation of the scintillation and Cherenkov components of the light with little losses, as described in Ref.~\cite{Kaptanoglu:2018sus}.
The use of light concentrators such as Winston cones covered with dichroic filters would provide a powerful way of separating the two light components while increasing light collection efficiency~\cite{Kaptanoglu:2019gtg}.

\subsubsection{Silicon Photomultiplier Modules}

The use of Silicon Photomultipliers, or SiPMs, has been increasingly widespread throughout the particle physics community~\cite{Simon:2018xzl}.
These solid-state single photon sensors have characteristics similar to those of photomultiplier tubes but with the advantages of much lower operating voltages and a very small size, usually on the mm-scale.
The latter is particularly interesting when trying to reconstruct light patterns since a surface covered with SiPMs could be considered as a highly pixelated sensor.
A tightly packed array of SiPMs would have the imaging capabilities of an LAPPD with added benefits of a less stringent voltage requirement and a more modular deployment.
However, it would also require more readout channels for the same area and would require tolerating a higher single photoelectron noise if operated at room temperature.

\section{Experimental Setup}
\label{sec:setup}

\subsection{Neutrinos from the Akkuyu Nuclear Power Plant}
\label{subsec:reactor}

Nuclear reactors are by far the most intense man-made source of antineutrinos.
Each fission reaction occurring in the reactor core releases on average 200~MeV of energy, several neutrons and two neutron-rich fission products.
These elements undergo a series of $\beta$-decays emitting on average of 6 electron antineutrinos (\ensuremath{\bar{\nu}_e}) in the process\footnote{In the remainder of this study, electron antineutrinos will be referred to as ``neutrinos'' for brevity.}.
The number, as well as the energy distribution of the neutrinos, is slightly dependent on the element at the origin of the fission chain ($^{235}$U, $^{238}$U, $^{239}$Pu or $^{241}$Pu).
During a reactor operation cycle, its fuel composition evolves and, as its uranium content decreases and more plutonium is produced, the rate of neutrinos and their energy distribution changes.
While a good understanding of this evolution is critical for precision measurements using reactor neutrinos, a constant fuel composition will be assumed in the rest of this study.

Using this information, the number of neutrinos emitted by a nuclear reactor core can be approximated to $2 \times 10^{20}$ \ensuremath{\bar{\nu}_e}.GW$_{\text{th}}$.s$^{-1}$, a rate directly dependent on the thermal power of the reactor.

The Akkuyu nuclear power plant, the construction of which started in 2018, is located in the Mersin province in Turkey.
It will be Turkey's first operational nuclear power plant and will consist of four identical VVER-1200 reactor units, each having an electrical power capacity of 1.2~GW$_{\text{e}}$.
Assuming a realistic thermal efficiency of 35\%, each reactor core will have a thermal capacity of about 3.4~GW$_{\text{th}}$ for a total output of 13.6~GW$_{\text{th}}$.
The Akkuyu reactors are Pressurized Water Reactors (PWR) and operate with fuel enriched in $^{235}$U to a level of about 5\%, classifying it as a Lowly Enriched Uranium, or LEU, fuel.
The summed energy spectrum of the neutrinos used in the remainder of this study was obtained from the fuel fractions of LEU from Ref.~\cite{Barna:2015rza}.

Historically, and due to its high cross section at MeV-scale energies, the Inverse Beta Decay (IBD) reaction has been widely used to detect electron antineutrinos.

This process, expressed as $\bar{\nu_{e}} + p \rightarrow e^{+} + n$, is a charged-current antineutrino scattering reaction occurring on a free proton and leads to the creation of a positron and a neutron.
The outgoing positron carries most of the incoming neutrino energy while the neutron is generated at energies ranging from $\sim$10-100~keV.
This reaction has a threshold of 1.8~MeV, meaning only a fraction of the neutrinos generated by fission products can be detected through IBD.
A detailed calculation of the differential IBD cross section can be found in Refs.~\cite{Vogel:1999zy,Strumia:2003zx}.

The high hydrogen content of water and organic liquids and plastics makes them attractive media to detect neutrinos through the IBD process.
In such materials, positrons deposit their energy and annihilate almost instantaneously while neutrons lose energy through elastic scattering, typically for a few 100's of microseconds, before being captured at thermal energies.
The coincidence of the two events is a key selection criteria for extracting neutrinos generated through IBD from other backgrounds.
In order to reduce the time difference between the prompt and delayed events (due to the positron and the neutron respectively), a dopant with a higher thermal neutron capture cross section can be utilized. 
Gadolinium with a neutron capture cross section of about 49,000 barns, is widely used by reactor neutrino experiments.
Upon capturing a neutron, it emits a gamma cascade with a total energy of about 8~MeV unlike the 2.2~MeV gamma originating from a capture on hydrogen, this is significantly above most backgrounds and drastically enhancing detection efficiency.  
Furthermore, its high cross section also shortens the neutron diffusion time, allowing a faster capture and yet further background rejection.
A loading of 0.1\% Gd by mass shortens the average neutron capture time from about 200~$\mu$s in an unloaded medium to about 30~$\mu$s.
Other neutron capture enhancing elements, such as lithium or chlorine, have also been used in previous experiments~\cite{Ashenfelter:2019iqj,Aharmim:2005gt}.

\subsection{Detector Design}
\label{subsec:detectors}

\begin{figure}[hbt!]
\centering 
\includegraphics[width=0.6\textwidth]{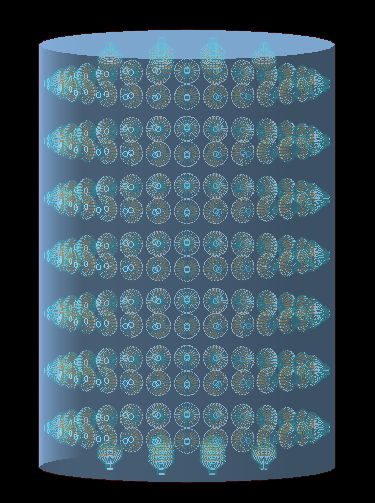}
\caption{Schematic view of the inner volume of the detector. The shields and vetos are not displayed.}
\label{fig:detector_geant4}
\end{figure}

For this study, a monolithic detector using a liquid detection medium is considered.
The detector consists of a 30-ton cylindrical volume, 4.3~m high and 3~m in diameter, containing Water-based Liquid Scintillator with a 10\% liquid scintillator content and a gadolinium loading of 0.1\% by mass.
A sketch of the detector and its photosensors is displayed in Figure~\ref{fig:detector_geant4}.

The choice of detection medium is motivated by the fact that this detector is intended as a cost-effective testbed for new technologies\footnote{Similar experiments with a water or liquid scintillator detector have been carried to completion and present little interest for the purpose of developing an R\&D program.}.
It is instrumented with 227~10-inch High Quantum Efficiency PMTs, placed on the walls of the cylindrical tank and facing the inside of the volume, providing a total photocoverage of around 30\%.

The walls and top cap of the detector are covered with Resistive Plate Chambers (RPCs) to tag charged particles, mainly atmospheric muons, entering the inner volume.
The detection efficiency of such gas detectors is known to reach values above 99\%~\cite{Sirunyan:2018fpa} making them a reliable and cost-efficient veto apparatus.
Further atmospheric muon tagging can be performed using the high amount of visible energy left by a muon track in the detector volume.
In order to reduce the external backgrounds from natural radioactivity or atmospheric muon spallation on the surrounding materials, the detector will be situated in a cavity and surrounded by a passive water shield with a thickness of 50~cm.

\begin{figure}[hbt!]
\centering 
\includegraphics[width=0.75\textwidth]{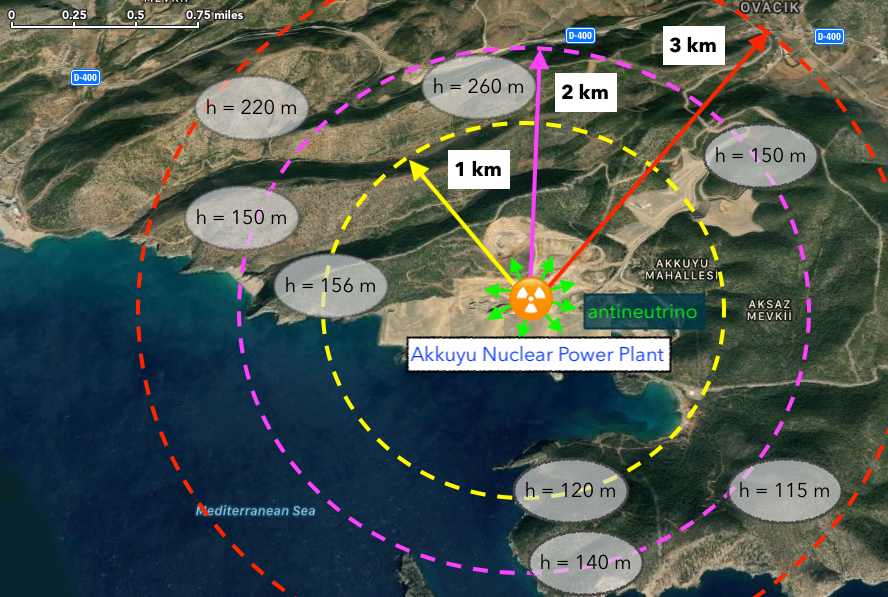}
\caption{Satellite image of the construction site of the future Akkuyu Nuclear Power Plant and its surrounding. Three isodistances of 1, 2 and 3~km from the expected reactors' locations are shown. Overburdens in term of rockbed thicknesses are also shown for several locations of interest around the power plant. Image extracted from Google Maps.}
\label{fig:akkuyu_top_view}
\end{figure}

In order to reduce the rate of atmospheric muons passing through the detector and its surroundings, the detector must be installed underground.
Fortunately, the hills surrounding the Akkuyu power plant are ideal locations, offering an ample overburden even at distances as close as 1~km from the plant, as shown in Figure~\ref{fig:akkuyu_top_view}.

For this study, we will assume our detector is installed in a cavern under the hill 1~km north of the plant offering a rock overburden of 250~meters, or 650 meter water equivalent (m.w.e) considering an average generic rock density of 2.6~g.cm$^{-3}$.
This location is optimal both in that it maximizes the neutrino flux, which follows a 1/D$^{2}$ behavior with D the distance from the detector to the reactors, and that it provides the most shielding overburden. 

A 30-ton detector located 1~km away from the Akkuyu power plant would see an interaction rate of about 200~neutrinos per day in the entirety of its volume before applying event selection cuts.
An estimation of the efficiency of these cuts is performed in Section~\ref{sec:simulations}.

\section{Simulation Studies}
\label{sec:simulations}

\subsection{Signal Simulation}
\label{signalsimulation}

To understand the detector response and assess its efficacy to detect reactor neutrinos, we simulated Inverse Beta Decay (IBD) events in the entire detector volume.
The conversion between deposited energy and the number of photoelectrons detected by the PMTs was established from simulations of electrons at the center of the detector. 
All simulations were performed using the GEANT4-based RAT-PAC simulation package\footnote{https://github.com/rat-pac/rat-pac}~\cite{Agostinelli:2002hh}.

When simulating WbLS loaded to a concentration of 10\%~liquid scintillator, we estimated its light yield to be 1000~photons per MeV - assuming a light yield of 10,000~photons per MeV for pure scintillator.
Along with scintillation, some light will also originate from Cherenkov emission. 
For a 1~MeV electron, we found that an average of 230 additional photons were generated through Cherenkov emission. 
As our simulated detector is using conventional PMTs as photosensors, distinction between these two sources of photons was beyond our capability and only their sum was used for the calorimetric measurement.
The use of fast photosensors, as discussed in Section~\ref{subsec:photosensors}, would allow the separation of these light components and provide more accurate energy and position reconstruction.

\begin{figure}[hbt!]
\centering
\begin{subfigure}[b]{1.0\textwidth}
\center{}
\includegraphics[scale=0.55]{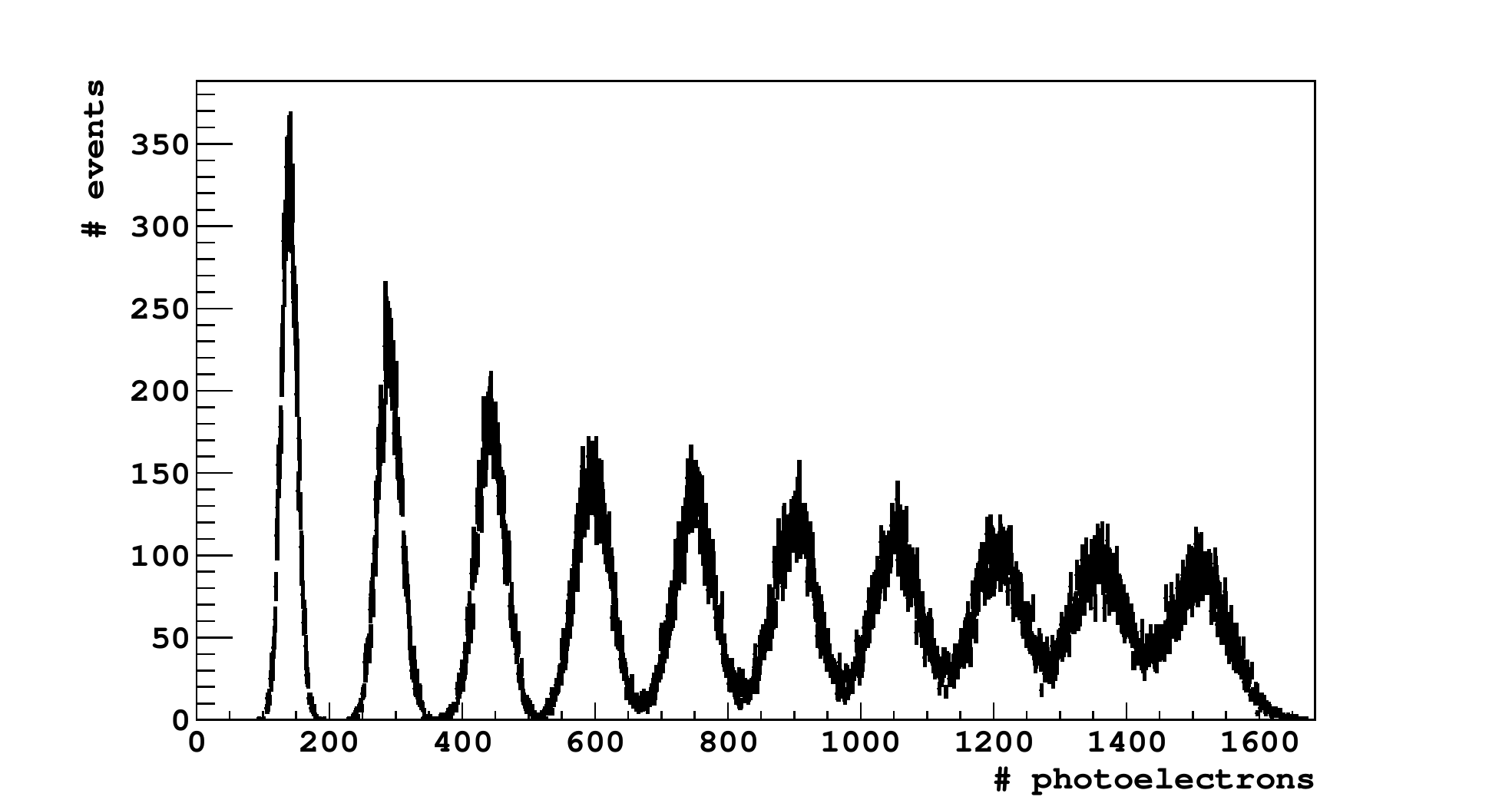}
\caption{}
\qquad{}
\end{subfigure}
\begin{subfigure}[b]{1.0\textwidth}
\center{}
\includegraphics[scale=0.55]{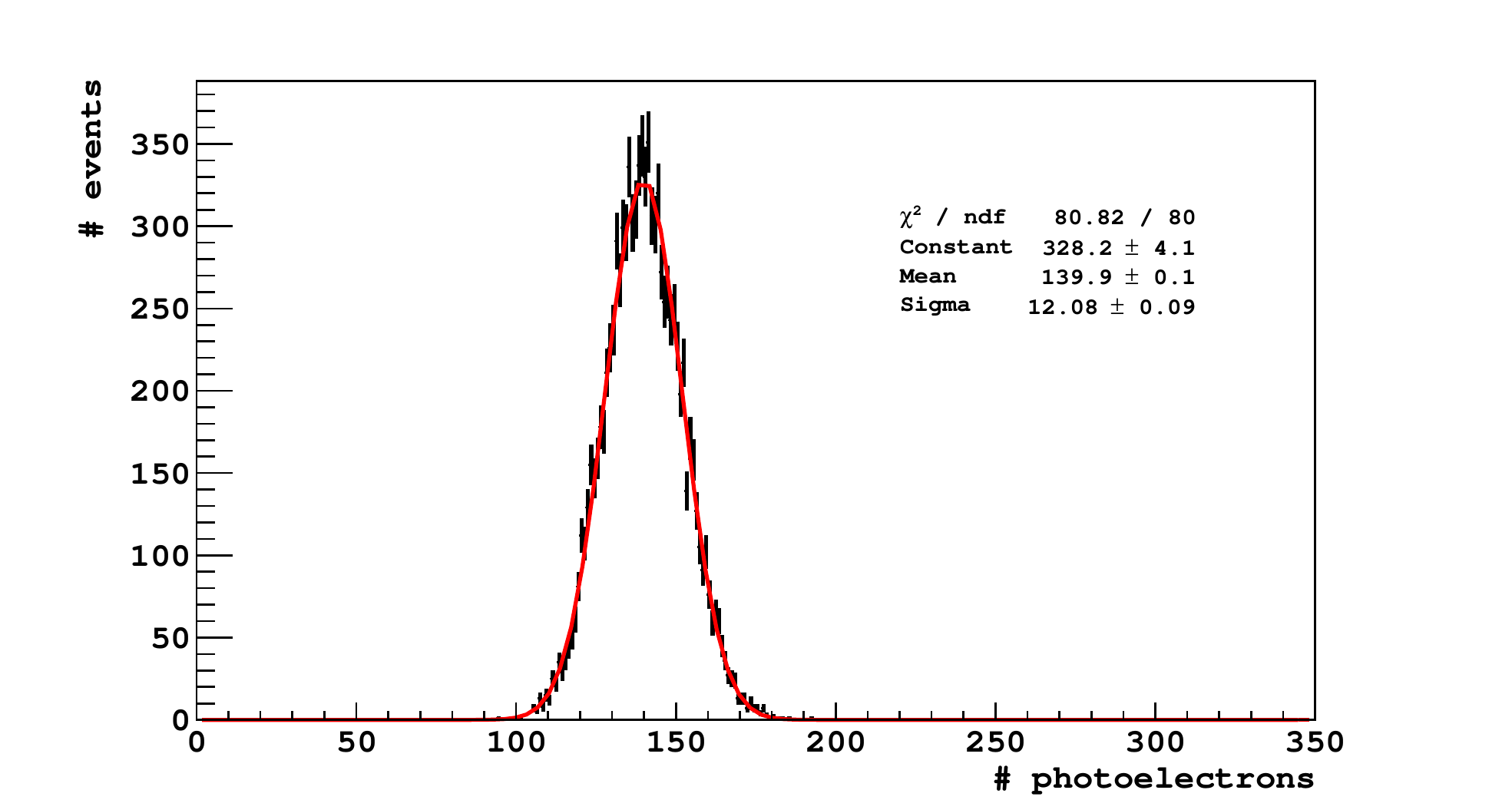}
\caption{}
\end{subfigure}

\caption {(a) Photoelectron distribution of electrons generated at the center of the detector with energies ranging from 1~to 10~MeV in increments of 1~MeV. (b) Photoelectron distribution of 1~MeV electrons generated at the center of the detector and fitted with a gaussian distribution.}
\label{fig:Ee_to_PE}
\end{figure}

As shown in Figure~\ref{fig:Ee_to_PE}, the resulting relation between electron energy and number of photoelectrons can be assumed to be linear at first order and 140~photoelectrons corresponds to 1~MeV.
A gaussian fit applied to the photoelectron distribution of 1~MeV electrons yields an energy resolution of 12\%.

\begin{figure}[hbt!]
\centering
\begin{subfigure}[b]{1.0\textwidth}
\center{}
\includegraphics[scale=0.55]{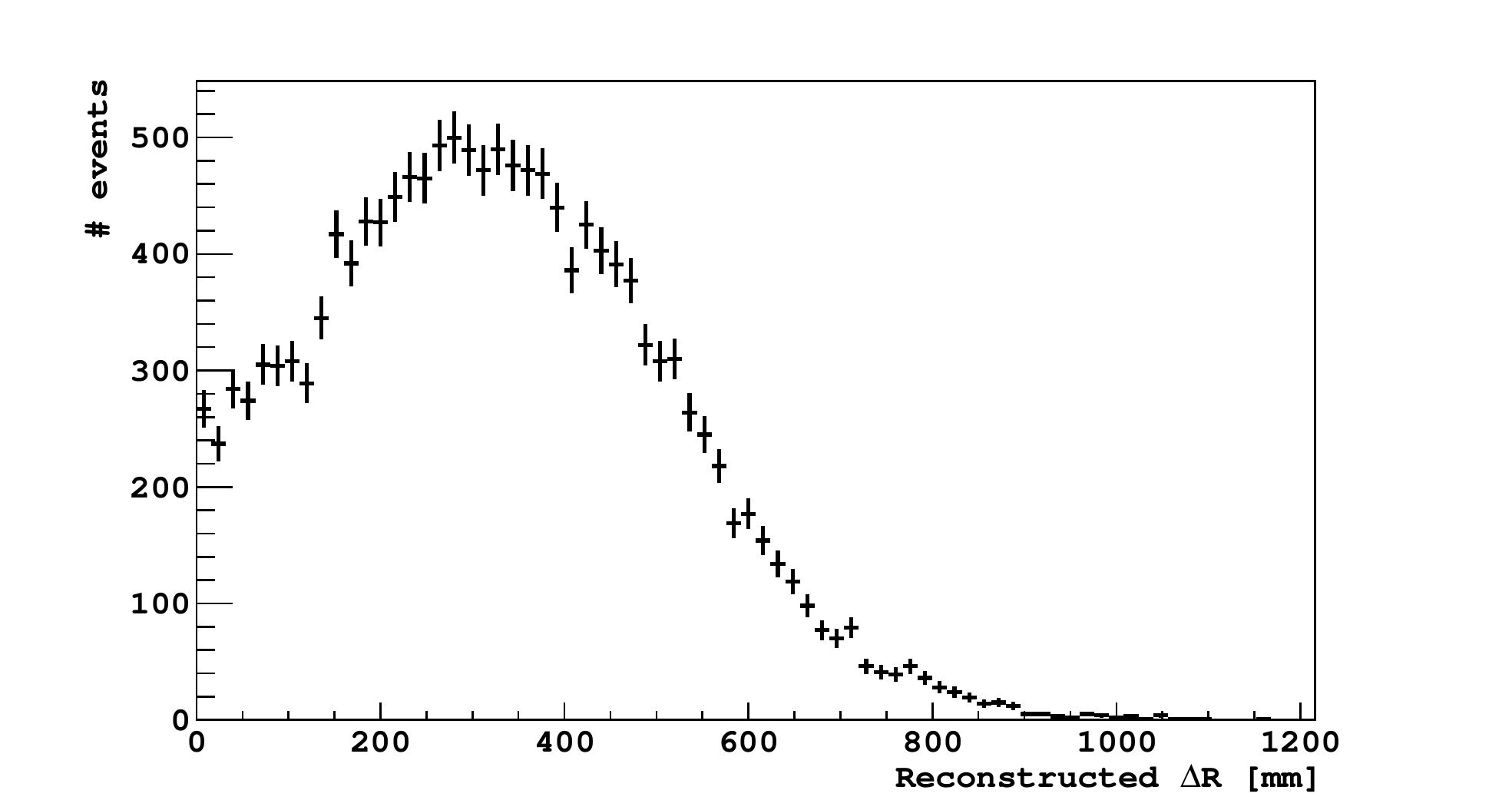}
\caption{}
\qquad    
\end{subfigure}
\begin{subfigure}[b]{1.0\textwidth}
\center{}
\includegraphics[scale=0.55]{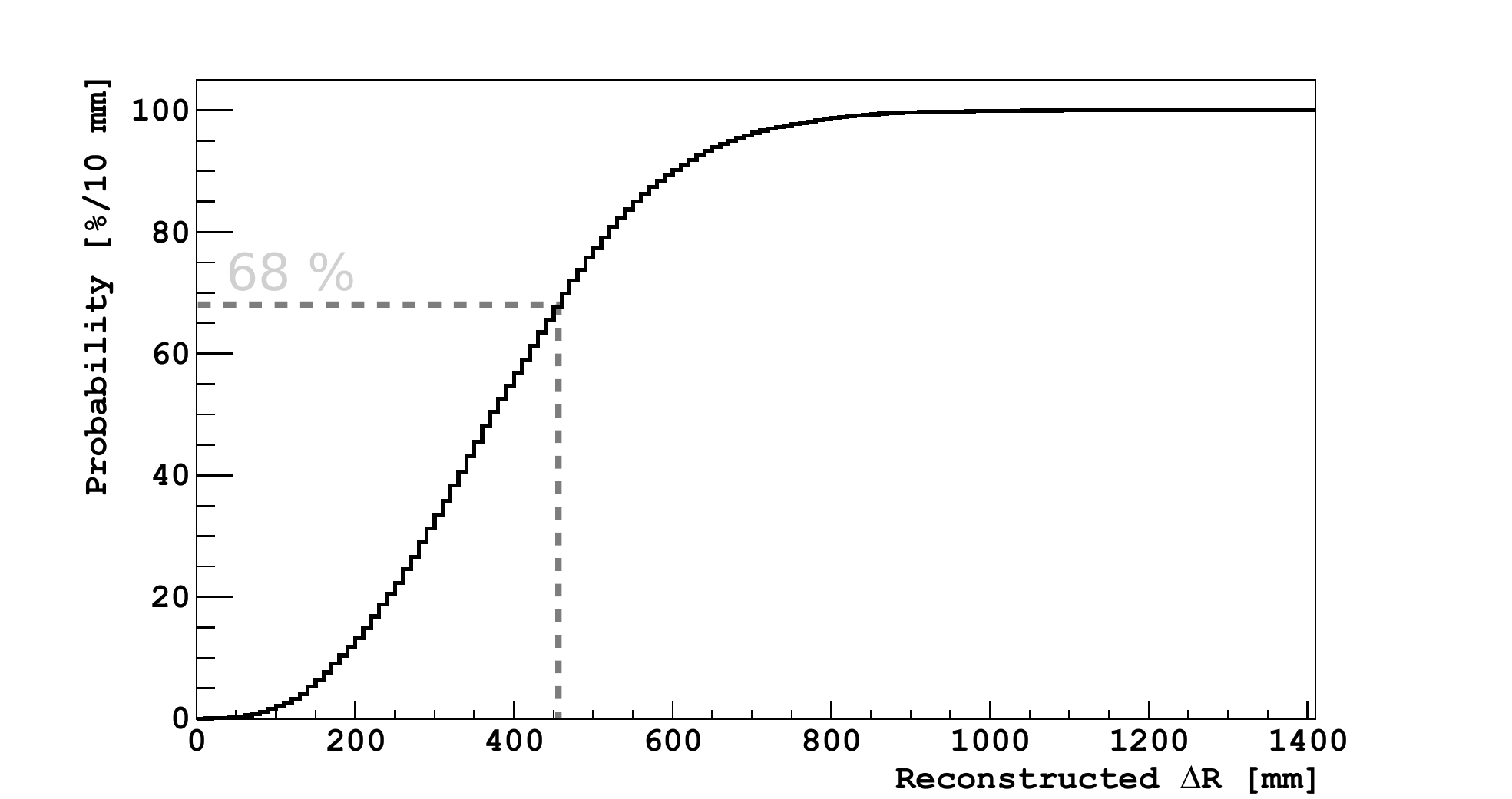}
\caption{}    
\end{subfigure}
\caption {Event-by-event $\Delta$R distribution (a) and cumulative $\Delta$R distribution (b) of 1~MeV electrons generated at the center of the detector. The vertex reconstruction is performed using the native RAT-PAC position reconstruction algorithm. 68\% of the interaction vertices are reconstructed within 45~cm of their true position.}
\label{fig:Ee_to_resolution}
\end{figure}

Simulating electrons also allowed us to assess the position reconstruction capabilities of the detector.
The event-by-event and cumulative distribution of $\Delta$R, the spatial separation between simulated and reconstructed vertices are shown in Figure~\ref{fig:Ee_to_resolution}.
The latter distribution yields a vertex reconstruction resolution of 45~cm at 1~MeV.

The energy and vertex resolutions are directly dependent on the number of detected photoelectrons.
The extrapolation at different energies can be obtained through the use of the following relations: 12\%/$\sqrt{E(MeV)}$ and 45~cm/$\sqrt{E(MeV)}$ for energy and vertex resolution respectively.

In order to simulate IBD events in the detector, we generated positron-neutron pairs with realistic energy and angular distributions obtained from Ref.~\cite{Strumia:2003zx}.
Since detection efficiency is expected to be largely suppressed outside of the volume surrounded by the PMTs, only interaction vertices within a cylinder of 1-m radius and 3.3-m height, centered in the detector, were considered in the following.
While 200 neutrino events are expected per day in the entire detector volume, limiting the analysis to this smaller inner volume reduces the event rate to 70~interactions per day.

\begin{figure}[hbt!]
\centering
\begin{subfigure}[b]{1.0\textwidth}
\center{}
\includegraphics[scale=0.47]{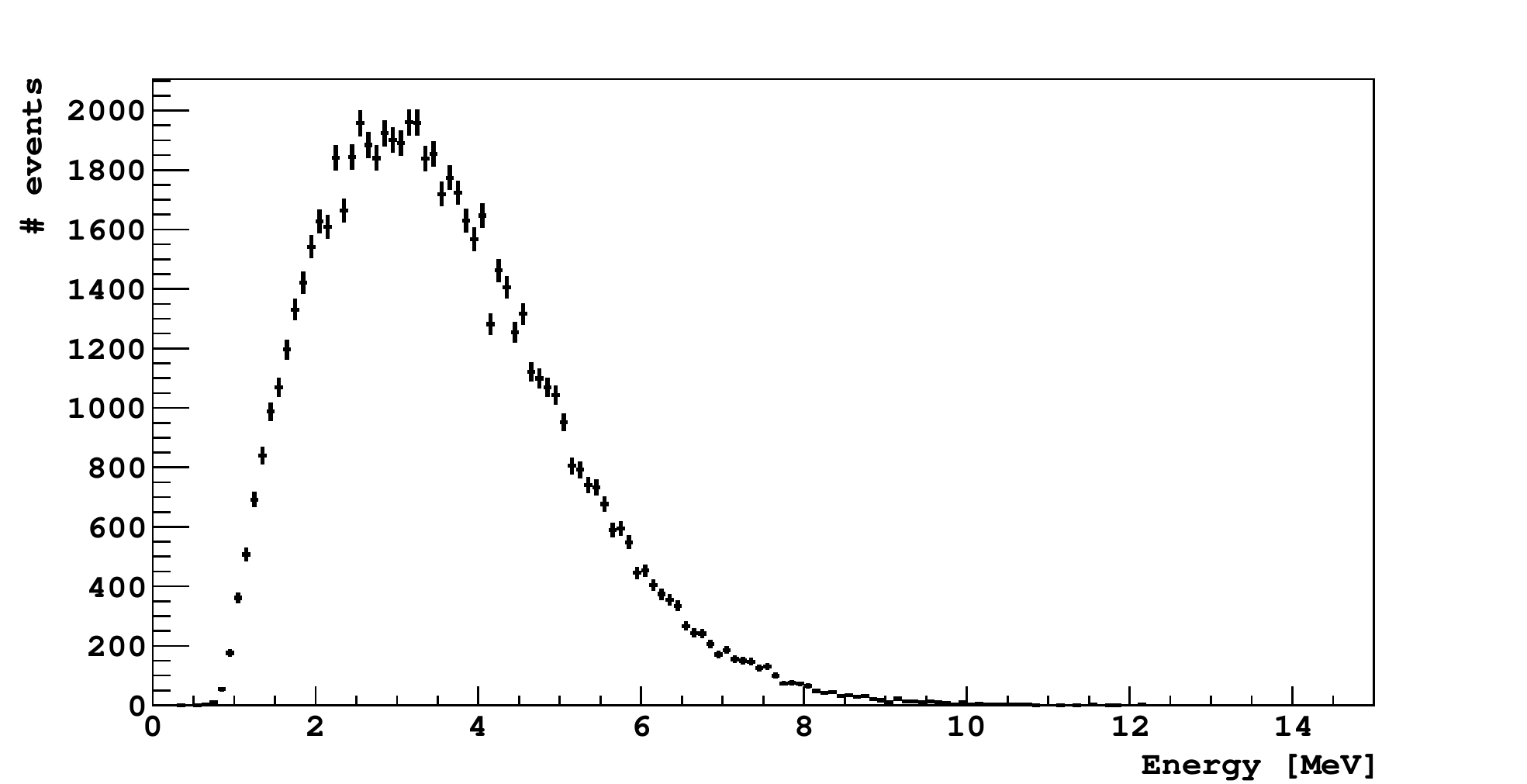}
\qquad
\caption{}    
\end{subfigure}
\begin{subfigure}[b]{1.0\textwidth}
\center{}
\includegraphics[scale=0.47]{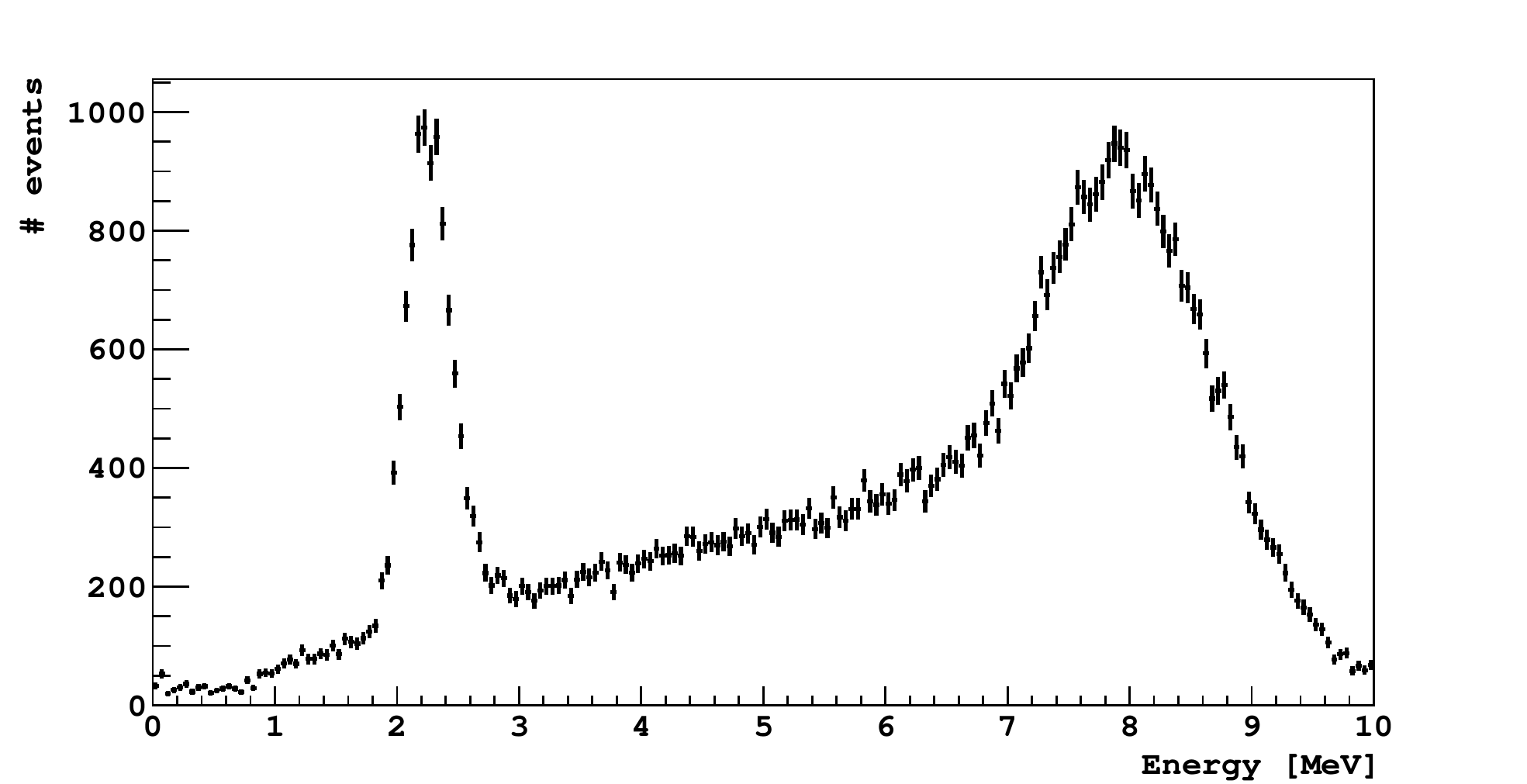}
\qquad    
\caption{}    
\end{subfigure}
\begin{subfigure}[b]{1.0\textwidth}
\center{}
\includegraphics[scale=0.47]{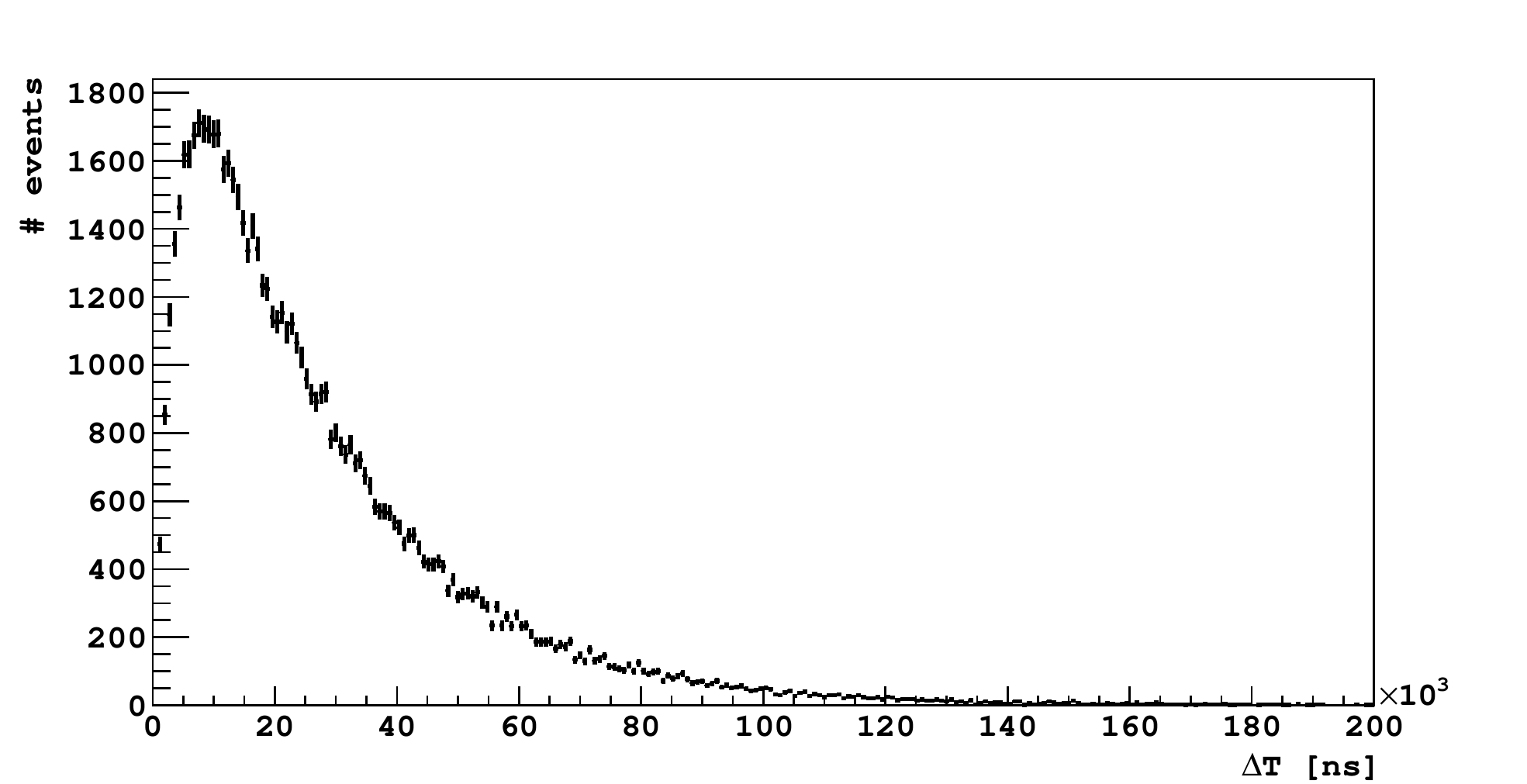}
\caption{}    
\end{subfigure}
\caption {Reconstructed prompt (a) and delayed (b) energy distributions, as well as time difference (c) for IBD events generated in the inner volume of the detector. }
\label{fig:IBD_Edp_DT}
\end{figure}

The photoelectron distributions from the prompt and delayed events, expressed in terms of electron energy using the conversion factor obtained from Figure~\ref{fig:Ee_to_PE}, as well as the time difference between the detection of the two events are shown in Figure~\ref{fig:IBD_Edp_DT}.
The prompt energy distribution corresponds to the visible energy deposited by positrons in the detector and is directly related to the neutrino energy at first order.
The delayed energy distribution exhibits two structures: the narrow 2.2~MeV peak from neutron captures on hydrogen and the broader peak from neutron captures on gadolinium centered around 8~MeV.
The tail of the latter at lower energies is due to gamma cascades where one or several gammas escape the detector volume.
The time difference between the prompt and delayed events follows an exponential distribution with a $\sim$24~$\mu$s mean decay time, as expected for a medium loaded with 0.1\% gadolinium.

\subsection{Signal Selection}
\label{signalselection}

Background rejection is critical for detecting reactor neutrinos.
Fortunately, the positron-neutron pair of IBD products provides a strong set of observables that can be used to suppress backgrounds events, mostly considered uncorrelated with each other.
Applying a selection cut on the time distribution helps discriminate IBD events from random background coincidences.
Selecting events whose prompt-delayed time difference falls between 0~and~100~$\mu$s, given the $\sim$24~$\mu$s mean capture time of neutrons in a 0.1\% Gd-loaded detection medium, allows us to maintain a signal purity of 98\% while discarding most accidental coincidences.
Similarly, selecting delayed events whose energy is reconstructed between 4~and 10~MeV is a strong method to eliminate backgrounds from natural radioactivity, such as $^{40}$K (1.461~MeV gamma ray) and $^{208}$Tl (2.614~MeV gamma ray), and better isolate neutron captures on gadolinium generating $\sim$8~MeV gamma cascades.
Such an energy cut yields a signal purity of 74\%.
Without applying an energy cut on the reconstructed prompt energy, to fully observe the spectrum of reactor neutrinos, the two aforementioned cuts yield a 73\% efficiency of isolating signal IBD pairs with a strong background discrimination.

After these cuts, about 50~neutrinos per day are expected to be efficiently reconstructed in a 30-ton WbLS detector deployed 1~km away from the Akkuyu NPP.

\section{Discussion}
\label{sec:discussions}

This proposed WbLS detector in Turkey aims to be an excellent modular test platform for the development of technology for future neutrino experiments around the globe. 
Such an effort would directly benefit future experiments planning on deploying WbLS as their detection medium such as WATCHMAN and THEIA.
The WATCHMAN experiment aims at installing a Gd-loaded water Cherenkov detector with a fiducial mass of 1~kton in the Boulby mine, UK, to monitor neutrinos from the Hartlepool reactor complex, located 25~km away.
While the base design consists of a Gd-loaded water medium, planned upgrades are in discussion and include the partial or complete deployment of WbLS in the detector volume. 
THEIA is a 25-100~kt neutrino detector project whose aim is to use WbLS as a detection medium in combination with state-of-the-art photon detection technologies such as the ones presented in Section~\ref{subsec:photosensors}.
Such a design would allow THEIA to perform world-leading measurement over a broad range of neutrino energies.
Both projects would highly benefit from prior smaller testbed experiments to provide valuable inputs in terms of technological R\&D and physics potential over a broad range of energies.

This WbLS detector would be complementary with the ANNIE experiment, a 26-ton Gd-loaded water Cherenkov detector installed on the Booster Neutrino Beam at Fermilab.
The ANNIE collaboration is aiming to deploy WbLS in a 1-ton vessel inside the detector volume within the next year and study its potential for high energy neutrino physics with a combination of conventional and fast photosensors.
By deploying a WbLS detector in the vicinity of a nuclear power plant, one can investigate its capabilities for low energy neutrino physics amidst higher backgrounds.
Such capabilities, enhanced by the use of fast photosensors, include, but are not limited to:
\begin{itemize}
    \item background rejection using Cherenkov and scintillation light separation,
    \item atmospheric muon tagging using scintillation calorimetry and Cherenkov light profile,
    \item directional reconstruction of Cherenkov rings from low energy neutrino-induced electron scattering.
\end{itemize}
The study of such background rejection and signal reconstruction techniques is of a particular interest for nuclear non-proliferation applications as it could increase the detection sensitivity of smaller and shallower neutrino detectors.

For this detector apparatus, we do not entirely rule out the possibility to use other detector media such as pure water, Gd-doped water or liquid scintillator. 
Water Cherenkov detectors are cost-efficient and easily scalable, but in the case of reactor monitoring they come with the disadvantage of a lower detection efficiency for low energy events. 
On the other hand, liquid scintillator detectors offer a lower detection threshold given their higher light yield but their size is limited by cost and light attenuation.
Since both technologies have been successfully used in the past to detect low energy neutrinos~\cite{DoubleChooz:2019qbj,Adey:2018zwh,Bak:2018ydk,Abe:2016nxk} their deployment as the single detection medium for this detector presents little interest.
However, a combination of the two, for instance WbLS in an outer volume and pure scintillator inside an inner volume, would be an option that could be investigated.
Such a scenario would provide crucial inputs for a THEIA-like detector aimed at studying neutrinoless double-beta decay.

In this study we consider a WbLS detector with a 10\% scintillator loading and a 0.1\% gadolinium doping.
The gamma cascade released by gadolinium de-excitation upon a neutron capture induces a spatially broad energy deposition in the vicinity of the capture location.
For applications where a precise vertex reconstruction is required, such as the reduction of high-intensity backgrounds or the reconstruction of the incoming neutrino flux direction~\cite{Apollonio:1999jg,Fischer:2015oma}, a point-like energy deposition is preferred.
For that purpose, several neutrino detectors are designed to use $^{6}$Li instead of Gd as their neutron capture enhancement isotope~\cite{Ashenfelter:2018zdm,Abreu:2017bpe}.
A neutron capture on $^{6}$Li leads to the creation of an alpha particle ($^{4}$He) and a tritium nucleus ($^{3}$H).
These two massive particles deposit their energy over such short tracks that they are usually considered point-like.
However, they do not generate Cherenkov light and their high energy deposition density suppresses the scintillation light output through the \textit{quenching} effect.
While the neutron capture reaction releases 4.78~MeV of energy, its light output is equivalent to an electron having an energy of about 0.5~MeV.
Such a low energy event would require a higher scintillator concentration in WbLS to be efficiently detected and reconstructed.
Li-doped WbLS would thus be worth investigating if the scintillator content could be increased up to 20\%.
In a detector such as the one presented in this study, this would lead to the detection of more than 100 photoelectrons per neutron capture on $^{6}$Li.

As previously stated, background mitigation is of the utmost importance for low energy reactor neutrino detection.
Internal and external gamma backgrounds, mostly caused by radioactive decays in the detector or its surroundings must be kept to a minimum.
This is achieved by careful screening and the use of low activity materials (acrylic, low radioactivity PMT glass, etc.), purifying the detection medium at the manufacturing stage or in-situ, and installing an adequate amount of shielding around the detector.
Backgrounds caused by atmospheric muons can be reduced by passive shielding and discarded using active vetos.
The most common and efficient way to reduce the flux of atmospheric muons through and around the detector is to build it underground, below a rock overburden.
Fortunately, the Akkuyu NPP is surrounded by several hills higher than 100~meters at distances as close as 1~km from the reactor cores, as shown in Figure~\ref{fig:akkuyu_top_view}.
Such an overburden significantly reduces the backgrounds caused by atmospheric muon interactions: (1) fast neutrons, generated by muon spallation on the surrounding materials and causing correlated background pairs (prompt from proton recoils and delayed from a neutron capture), (2) cosmogenics, radioactive isotopes generated through spallation and susceptible to emit neutrons upon decay, and (3) stopping muons, generating a prompt signal as they enter the detector and a delayed signal as they decay into Michel electrons.
Dedicated studies performed in Refs.~\cite{Abe:2009aa} and~\cite{deKerret:2018fqd} provide a very useful calculation of such backgrounds for different muon fluxes.
Such an extrapolation should be performed upon estimating or detecting the muon flux at the proposed location for the detector presented in this study.
Additional studies presented in Refs.~\cite{Li:2015kpa} and~\cite{Li:2015lxa} showed the efficacy of background rejection cuts based on reconstructed muon profiles, an observable enhanced by the use of fast photosensors.

With 50~neutrino events reconstructed per day, this WbLS detector will have the capability to perform a real-time monitoring of the Akkuyu NPP thermal power at a distance considered non-intrusive for the plant operation, a feature of interest for nuclear safeguard applications~\cite{IAEA_report}.
In order to fully exploit its potential as a demonstrator for a safe, reliable and cost-effective nuclear non-proliferation neutrino detector, a strong collaborative effort with the power plant operator would be beneficial.
This partnership would include sharing information such as the daily thermal power recorded by the operator and the estimated dates of refueling.
Other, more sensitive information such as fuel content and fuel assembly geometry could be shared with selected members of the detector collaboration.
We foresee that this joint effort will strengthen the ties between the fundamental particle physics and the nuclear energy communities.

\section{Conclusion}
\label{sec:conclusion}

In this paper, we present the first simulation results of a proposed medium-scale Water-based Liquid Scintillator (WbLS) detector, designed to become a new technology testbed for neutrino studies in Turkey near the country's first nuclear power plant, the Akkuyu Nuclear Power Plant (Akkuyu NPP). 
The proposed detector is a 30-ton WbLS detector in the vicinity of the Akkuyu NPP. 
At a distance of 1~km and within the entire detector volume, 200~neutrino interactions are expected per day.
After applying signal selection and background rejection cuts to reconstruct reactor neutrinos with a higher purity, an estimated rate of 50~neutrino events per day is expected. 
With such a rate, this detector is well suited to serve as an effective monitoring tool of the Akkuyu NPP, and also as a detector R\&D testbed for the low energy program of future large-scale international neutrino experiments such as WATCHMAN and THEIA. Such a project would be a valuable opportunity for the neutrino and nuclear physics communities in Turkey to perform world-leading R\&D with strong ties to the international neutrino physics community.

\section{Acknowledgements}
\label{sec:acknowledgements}

The authors would like to thank Leon Pickard and Marcus O'Flaherty for their prompt and careful reading of this paper.
V. Fischer is supported by the DOE National Nuclear Security Administration through the Nuclear Science and Security Consortium under award number DE-NA0003180 and the U.S. Department of Energy (DOE) Office of Science under award number DE-SC0009999. 
E. Tiras is supported by the U.S. Department of Energy, Office of Science, Office of High Energy Physics, Intensity Frontier Experimental Research program under Award Number DE-SC0017946 and Prof. Wetstein's startup fund at Iowa State University. 


\bibliographystyle{unsrt}
\bibliography{bibliography.bib}

\begin{thebibliography}{10}
\expandafter\ifx\csname url\endcsname\relax
  \def\url#1{\texttt{#1}}\fi
\expandafter\ifx\csname urlprefix\endcsname\relax\def\urlprefix{URL }\fi
\expandafter\ifx\csname href\endcsname\relax
  \def\href#1#2{#2} \def\path#1{#1}\fi

\bibitem{Ozturk:2016tjn}
S.~Ozturk, A.~Adiguzel, V.~E. Ozcan, N.~G. Unel, {Monitoring Akkuyu Nuclear
  Reactor Using Antineutrino Flux Measurement}, Turk. J. Phys. 41~(1) (2017)
  41--46.
\newblock \href {http://arxiv.org/abs/1602.04646} {\path{arXiv:1602.04646}},
  \href {http://dx.doi.org/10.3906/fiz-1604-20}
  {\path{doi:10.3906/fiz-1604-20}}.

\bibitem{Kandemir:2019dfs}
M.~Kandemir, A.~Cakir, {A reactor antineutrino detector based on hexagonal
  scintillator bars}, Nucl. Instrum. Meth. A953 (2020) 163251.
\newblock \href {http://arxiv.org/abs/1908.08117} {\path{arXiv:1908.08117}},
  \href {http://dx.doi.org/10.1016/j.nima.2019.163251}
  {\path{doi:10.1016/j.nima.2019.163251}}.

\bibitem{Ozturk:2019bul}
S.~Ozturk, {Nuclear Reactor Monitoring with Gadolinium-Loaded Plastic
  Scintillator Modules, }\href {http://arxiv.org/abs/1906.01944}
  {\path{arXiv:1906.01944}}, \href
  {http://dx.doi.org/10.1016/j.nima.2019.163314}
  {\path{doi:10.1016/j.nima.2019.163314}}.

\bibitem{WBLS1}
M.~Yeh, et~al., {A New Water-based Liquid Scintillator and Potential
  Applications}, Nucl. Instrum. Meth. A660 (2011) 51.

\bibitem{Adams:2016tfm}
B.~W. Adams, et~al., {A Brief Technical History of the Large-Area Picosecond
  Photodetector (LAPPD) Collaboration}, Submitted to: JINST,\href
  {http://arxiv.org/abs/1603.01843} {\path{arXiv:1603.01843}}.

\bibitem{Kaptanoglu:2018sus}
T.~Kaptanoglu, M.~Luo, J.~Klein, {Cherenkov and Scintillation Light Separation
  Using Wavelength in LAB Based Liquid Scintillator}, JINST 14~(05) (2019)
  T05001.
\newblock \href {http://arxiv.org/abs/1811.11587} {\path{arXiv:1811.11587}},
  \href {http://dx.doi.org/10.1088/1748-0221/14/05/T05001}
  {\path{doi:10.1088/1748-0221/14/05/T05001}}.

\bibitem{Kaptanoglu:2019gtg}
T.~Kaptanoglu, M.~Luo, B.~Land, A.~Bacon, J.~Klein, {The Dichroicon: Spectral
  Photon Sorting For Large-Scale Cherenkov and Scintillation Detectors, }\href
  {http://arxiv.org/abs/1912.10333} {\path{arXiv:1912.10333}}.

\bibitem{Askins:2015bmb}
M.~Askins, et~al., {The Physics and Nuclear Nonproliferation Goals of WATCHMAN:
  A WAter CHerenkov Monitor for ANtineutrinos, }\href
  {http://arxiv.org/abs/1502.01132} {\path{arXiv:1502.01132}}.

\bibitem{Askins:2019oqj}
M.~Askins, et~al., {Theia: An advanced optical neutrino detector, }\href
  {http://arxiv.org/abs/1911.03501} {\path{arXiv:1911.03501}}.

\bibitem{Beacom:2003nk}
J.~F. Beacom, M.~R. Vagins, {GADZOOKS! Anti-neutrino spectroscopy with large
  water Cherenkov detectors}, Phys. Rev. Lett. 93 (2004) 171101.
\newblock \href {http://arxiv.org/abs/hep-ph/0309300}
  {\path{arXiv:hep-ph/0309300}}, \href
  {http://dx.doi.org/10.1103/PhysRevLett.93.171101}
  {\path{doi:10.1103/PhysRevLett.93.171101}}.

\bibitem{Back:2017kfo}
A.~R. Back, et~al., {Accelerator Neutrino Neutron Interaction Experiment
  (ANNIE): Preliminary Results and Physics Phase Proposal, }\href
  {http://arxiv.org/abs/1707.08222} {\path{arXiv:1707.08222}}.

\bibitem{Lyashenko:2019tdj}
A.~V. Lyashenko, et~al., {Performance of Large Area Picosecond Photo-Detectors
  (LAPPD), }\href {http://arxiv.org/abs/1909.10399} {\path{arXiv:1909.10399}},
  \href {http://dx.doi.org/10.1016/j.nima.2019.162834}
  {\path{doi:10.1016/j.nima.2019.162834}}.

\bibitem{Tiras2019}
E.~Tiras, {Detector R\&D for ANNIE and Future Neutrino Experiments}, in:
  {APS-DPF Meeting, Boston, Massachusetts, July 29-August 2, 2019}.
\newblock \href {http://arxiv.org/abs/1910.08715} {\path{arXiv:1910.08715}}.

\bibitem{Simon:2018xzl}
F.~Simon, {Silicon Photomultipliers in Particle and Nuclear Physics}, Nucl.
  Instrum. Meth. A926 (2019) 85--100.
\newblock \href {http://arxiv.org/abs/1811.03877} {\path{arXiv:1811.03877}},
  \href {http://dx.doi.org/10.1016/j.nima.2018.11.042}
  {\path{doi:10.1016/j.nima.2018.11.042}}.

\bibitem{Barna:2015rza}
A.~Barna, S.~Dye, {Global Antineutrino Modeling: A Web Application, }\href
  {http://arxiv.org/abs/1510.05633} {\path{arXiv:1510.05633}}.

\bibitem{Vogel:1999zy}
P.~Vogel, J.~F. Beacom, {Angular distribution of neutron inverse beta decay,
  anti-neutrino(e) + p---\ e+ + n}, Phys. Rev. D60 (1999) 053003.
\newblock \href {http://arxiv.org/abs/hep-ph/9903554}
  {\path{arXiv:hep-ph/9903554}}, \href
  {http://dx.doi.org/10.1103/PhysRevD.60.053003}
  {\path{doi:10.1103/PhysRevD.60.053003}}.

\bibitem{Strumia:2003zx}
A.~Strumia, F.~Vissani, {Precise quasielastic neutrino/nucleon cross-section},
  Phys. Lett. B564 (2003) 42--54.
\newblock \href {http://arxiv.org/abs/astro-ph/0302055}
  {\path{arXiv:astro-ph/0302055}}, \href
  {http://dx.doi.org/10.1016/S0370-2693(03)00616-6}
  {\path{doi:10.1016/S0370-2693(03)00616-6}}.

\bibitem{Ashenfelter:2019iqj}
J.~Ashenfelter, et~al., {Lithium-loaded Liquid Scintillator Production for the
  PROSPECT experiment}, JINST 14~(03) (2019) P03026.
\newblock \href {http://arxiv.org/abs/1901.05569} {\path{arXiv:1901.05569}},
  \href {http://dx.doi.org/10.1088/1748-0221/14/03/P03026}
  {\path{doi:10.1088/1748-0221/14/03/P03026}}.

\bibitem{Aharmim:2005gt}
B.~Aharmim, et~al., {Electron energy spectra, fluxes, and day-night asymmetries
  of B-8 solar neutrinos from measurements with NaCl dissolved in the
  heavy-water detector at the Sudbury Neutrino Observatory}, Phys. Rev. C72
  (2005) 055502.
\newblock \href {http://arxiv.org/abs/nucl-ex/0502021}
  {\path{arXiv:nucl-ex/0502021}}, \href
  {http://dx.doi.org/10.1103/PhysRevC.72.055502}
  {\path{doi:10.1103/PhysRevC.72.055502}}.

\bibitem{Sirunyan:2018fpa}
A.~M. Sirunyan, et~al., {Performance of the CMS muon detector and muon
  reconstruction with proton-proton collisions at $\sqrt{s}=$ 13 TeV}, JINST
  13~(06) (2018) P06015.
\newblock \href {http://arxiv.org/abs/1804.04528} {\path{arXiv:1804.04528}},
  \href {http://dx.doi.org/10.1088/1748-0221/13/06/P06015}
  {\path{doi:10.1088/1748-0221/13/06/P06015}}.

\bibitem{Agostinelli:2002hh}
S.~Agostinelli, et~al., {GEANT4: A Simulation toolkit}, Nucl. Instrum. Meth.
  A506 (2003) 250--303.
\newblock \href {http://dx.doi.org/10.1016/S0168-9002(03)01368-8}
  {\path{doi:10.1016/S0168-9002(03)01368-8}}.

\bibitem{DoubleChooz:2019qbj}
H.~de~Kerret, et~al., {First Double Chooz $\mathbf{\theta_{13}}$ Measurement
  via Total Neutron Capture Detection, }\href {http://arxiv.org/abs/1901.09445}
  {\path{arXiv:1901.09445}}.

\bibitem{Adey:2018zwh}
D.~Adey, et~al., {Measurement of the Electron Antineutrino Oscillation with
  1958 Days of Operation at Daya Bay}, Phys. Rev. Lett. 121~(24) (2018) 241805.
\newblock \href {http://arxiv.org/abs/1809.02261} {\path{arXiv:1809.02261}},
  \href {http://dx.doi.org/10.1103/PhysRevLett.121.241805}
  {\path{doi:10.1103/PhysRevLett.121.241805}}.

\bibitem{Bak:2018ydk}
G.~Bak, et~al., {Measurement of Reactor Antineutrino Oscillation Amplitude and
  Frequency at RENO}, Phys. Rev. Lett. 121~(20) (2018) 201801.
\newblock \href {http://arxiv.org/abs/1806.00248} {\path{arXiv:1806.00248}},
  \href {http://dx.doi.org/10.1103/PhysRevLett.121.201801}
  {\path{doi:10.1103/PhysRevLett.121.201801}}.

\bibitem{Abe:2016nxk}
K.~Abe, et~al., {Solar Neutrino Measurements in Super-Kamiokande-IV}, Phys.
  Rev. D94~(5) (2016) 052010.
\newblock \href {http://arxiv.org/abs/1606.07538} {\path{arXiv:1606.07538}},
  \href {http://dx.doi.org/10.1103/PhysRevD.94.052010}
  {\path{doi:10.1103/PhysRevD.94.052010}}.

\bibitem{Apollonio:1999jg}
M.~Apollonio, et~al., {Determination of neutrino incoming direction in the
  CHOOZ experiment and supernova explosion location by scintillator detectors},
  Phys. Rev. D61 (2000) 012001.
\newblock \href {http://arxiv.org/abs/hep-ex/9906011}
  {\path{arXiv:hep-ex/9906011}}, \href
  {http://dx.doi.org/10.1103/PhysRevD.61.012001}
  {\path{doi:10.1103/PhysRevD.61.012001}}.

\bibitem{Fischer:2015oma}
V.~Fischer, et~al., {Prompt directional detection of galactic supernova by
  combining large liquid scintillator neutrino detectors, }JCAP1508,032(2015).
\newblock \href {http://arxiv.org/abs/1504.05466} {\path{arXiv:1504.05466}},
  \href {http://dx.doi.org/10.1088/1475-7516/2015/08/032}
  {\path{doi:10.1088/1475-7516/2015/08/032}}.

\bibitem{Ashenfelter:2018zdm}
J.~Ashenfelter, et~al., {The PROSPECT Reactor Antineutrino Experiment}, Nucl.
  Instrum. Meth. A922 (2019) 287--309.
\newblock \href {http://arxiv.org/abs/1808.00097} {\path{arXiv:1808.00097}},
  \href {http://dx.doi.org/10.1016/j.nima.2018.12.079}
  {\path{doi:10.1016/j.nima.2018.12.079}}.

\bibitem{Abreu:2017bpe}
Y.~Abreu, et~al., {A novel segmented-scintillator antineutrino detector}, JINST
  12~(04) (2017) P04024.
\newblock \href {http://arxiv.org/abs/1703.01683} {\path{arXiv:1703.01683}},
  \href {http://dx.doi.org/10.1088/1748-0221/12/04/P04024}
  {\path{doi:10.1088/1748-0221/12/04/P04024}}.

\bibitem{Abe:2009aa}
S.~Abe, et~al., {Production of Radioactive Isotopes through Cosmic Muon
  Spallation in KamLAND}, Phys. Rev. C81 (2010) 025807.
\newblock \href {http://arxiv.org/abs/0907.0066} {\path{arXiv:0907.0066}},
  \href {http://dx.doi.org/10.1103/PhysRevC.81.025807}
  {\path{doi:10.1103/PhysRevC.81.025807}}.

\bibitem{deKerret:2018fqd}
H.~de~Kerret, et~al., {Yields and production rates of cosmogenic $^9$Li and
  $^8$He measured with the Double Chooz near and far detectors}, JHEP 11 (2018)
  053.
\newblock \href {http://arxiv.org/abs/1802.08048} {\path{arXiv:1802.08048}},
  \href {http://dx.doi.org/10.1007/JHEP11(2018)053}
  {\path{doi:10.1007/JHEP11(2018)053}}.

\bibitem{Li:2015kpa}
S.~W. Li, J.~F. Beacom, {Spallation Backgrounds in Super-Kamiokande Are Made in
  Muon-Induced Showers}, Phys. Rev. D91~(10) (2015) 105005.
\newblock \href {http://arxiv.org/abs/1503.04823} {\path{arXiv:1503.04823}},
  \href {http://dx.doi.org/10.1103/PhysRevD.91.105005}
  {\path{doi:10.1103/PhysRevD.91.105005}}.

\bibitem{Li:2015lxa}
S.~W. Li, J.~F. Beacom, {Tagging Spallation Backgrounds with Showers in
  Water-Cherenkov Detectors}, Phys. Rev. D92~(10) (2015) 105033.
\newblock \href {http://arxiv.org/abs/1508.05389} {\path{arXiv:1508.05389}},
  \href {http://dx.doi.org/10.1103/PhysRevD.92.105033}
  {\path{doi:10.1103/PhysRevD.92.105033}}.

\bibitem{IAEA_report}
IAEA, Final report: Focused workshop on antineutrino detection for safeguards
  applications report str-361, Tech. rep. (2008).

\end{thebibliography}

\end{document}